# Diffraction Influence on the Field of View and Resolution of Three-Dimensional Integral Imaging

Zhila Esna Ashari, Zahra Kavehvash, and Khashayar Mehrany

***Abstract*—The influence of the diffraction limit on the field of view of three-dimensional integral imaging (InI) systems is estimated by calculating the resolution of the InI system along arbitrarily tilted directions. The deteriorating effects of diffraction on the resolution are quantified in this manner. Two different three-dimensional scenes are recorded by real/virtual and focused imaging modes. The recorded scenes are reconstructed at different tilted planes and the obtained results for the resolution and field of view of the system are verified. It is shown that the diffraction effects severely affect the resolution of InI in the real/virtual mode when the tilted angle of viewing is increased. It is also shown that the resolution of InI in the focused mode is more robust to the unwanted effects of diffraction even though it is much lower than the resolution of InI in the real/virtual mode.**

## I. Introduction

THANKS TO the progression of technology in making high-resolution imaging sensors, display devices, and high quality yet cost-effective micro-lens arrays, there has been a resurgence of interest in the more-than-a-century old technique of Lippmann for three-dimensional (3D) imaging, which had been referred to as integral photography [1]. Taking on the new guise of integral imaging (InI), the very same principles have been extensively developed, refined, and applied in many directions [2]-[6]. Nevertheless, the field of view (FOV) and the achievable resolution of InI in its most basic form, i.e. when the many perspectives of the 3D scene of interest are directly captured via a lens array and when the 3D scene is optically reconstructed via a display device, are still open to debate [6]-[10]. The ongoing dispute over the FOV and the resolution of InI, which is very well reflected in the number of publications dedicated to the issue, is fueled by the fact that the performance of InI is limited by quite a large number of factors, viz. the size of lenslets in the lens array [11], lens pitch, the number of pixels allocated to each lenslet, display device resolution [12], [13], the position of the viewer [14], etc.

Generally speaking, the performance of InI systems is limited either by the optoelectronic devices such as CMOS, CCD, and LCDs needed for image recording and reconstruction stages, or by the lenslets in the lens array [12], [13]. Since technology advancement gives us hope to lay hands on better optoelectronic devices that can accommodate higher resolution and better performance, the fundamental limiting factor of the resolution in InI systems seems to stem from the restrictions inflicted by the optical response of the lens array. This is particularly true for InI systems that employ microlens array whose diffraction and defocus aberrations are by no means negligible [15]. For this reason, the focus of this work is on considering the unwanted effects of diffraction and defocus aberrations on the performance of lenslet-based InI systems. To this end, the principles of the wave optics are employed to assess the resolution of the typical InI systems not only in the lateral and axial directions but also along arbitrary directions. By doing so, the fundamental limit of the FOV in typical InI systems is rigorously quantified. It is worth noting that even though the free view reconstruction of 3D images along arbitrary directions by using the rules of geometrical optics have been already reported [16]-[18], simulation of optical reconstruction of 3D images by considering the effects of diffraction and defocus aberrations has to the best of our knowledge not been performed before. The only previous attempt to study the unwanted effects of diffraction on the performance of InI systems was limited to specific imaging directions [19].

The rest of the paper is organized as follows. First, an optical point source placed along the central axis of the lens array is visualized by considering the effects of diffraction. Using the Gaussian wave approximation, the volumetric representation of the visualized point source is obtained. The extent of the visualized point source at arbitrarily tilted planes is extracted and the resolution of the system is derived. The obtained resolutions justify the observed results of the next section, wherein the tilted views of two different three-dimensional scenes are reconstructed. Mathematical equations needed for the free view reconstruction of three-dimensional images are provided. The effects of diffraction and defocus aberration are taken into account. Finally, the conclusions are made.

Z. Esna Ashari, Z. Kavehvash and K. Mehrany are with the Department of Electrical Engineering, Sharif University of Technology, Tehran, Iran (e-mail: esnaashari@alum.sharif.edu; kavehvash@sharif.edu; mehrany@sharif.edu ).

## II. Extraction of InI resolution along tilted directions: Gaussian beam approximation

In this section, the achievable resolution of direct optical reconstruction in a typical InI system is estimated by taking the effects of diffraction into account. It is assumed that the InI system is composed of an $m \times n$ lens-array placed at distance $g$ from an ideal display device whose resolution is high enough to not restrain the overall resolution of the system. The lateral and longitudinal coordinates of the three-dimensional image space are designated by $x$,$y$, and $z$, respectively. The ($x$,$z$) cross section of the considered system is schematically shown in Fig. 1. The lens pitches along the $x$ and $y$ directions are represented by $s_x$, and $s_y$ respectively. It is also assumed that the cross section of each lenslet in the lens-array is an ellipse whose major and minor radii are equal to the lens pitches $s_x$ and $s_y$.

To estimate the resolution of such an InI system, an optical point source should be visualized. The extent of the optically visualized point source along any arbitrary direction in the three-dimensional image space is a good estimate of the resolution of the system along that arbitrary direction at the location of the point source. For simplicity's sake, the optical point source to be optically captured by the considered InI system is assumed to be located on the longitudinal axis at $z=D$ from the lens-array. To obtain the resolution of the system along arbitrary tilted directions, the size of the optical visualization of this point source should be extracted in the tilted plane. xWithouty losing generality, the tilt angles of the tilted directions are assumed $\theta$, and $\theta$, with respect to the $x$, and $y$ coordinates respectively. The lateral coordinates of the tilted plane are designated by $x_\theta$, and $y_\theta$, respectively. The normal to the tilted plane direction is designated by $z_\theta$. The following relations are held between the coordinates of the three-dimensional image space and the coordinates of the titled plane:

$$x = x_\theta \cos\theta_x$$

$$y = y_\theta \cos\theta_y \tag{1}$$

$$z = x_\theta \sin\theta_x + y_\theta \sin\theta_y + D$$

According to the Gaussian beam approximation [20], the volumetric image of the point source formed by the central lenslet in the lens array, when $m$ and $n$ are even numbers, can be reasonably approximated by a Gaussian beam:

$$O_{m/2,n/2}(x,y,z) = \frac{2P}{\pi w_x w_y} e^{-2\frac{x^2}{w_x^2}} e^{-2\frac{y^2}{w_y^2}} \tag{2}$$

where $P$ is the total power of the beam and $w_x$, and $w_y$ are the waists of the Gaussian beam along the $x$, and $y$, directions and can be written as [20]:

$$w_x(z) = w_{0x}\left[1+4\left(\frac{z-z_i}{b_x}\right)^2\right]^{1/2}$$

$$w_y(z) = w_{0y}\left[1+4\left(\frac{z-z_i}{b_y}\right)^2\right]^{1/2} \tag{3}$$

In these expressions, $z_i$ is the focused plane satisfying the lens law:

$$\frac{1}{z_i} + \frac{1}{g} = \frac{1}{f} \tag{4}$$

g is the distance between the object and the lens. And $b_x$ and $b_y$ are the Rayleigh range:

$$b_x = \frac{\pi w_{0x}^2}{2\lambda}$$

$$b_y = \frac{\pi w_{0y}^2}{2\lambda} \tag{5}$$

and

$$w_{0x} = 2.44\frac{\lambda z_i}{s_x}$$

$$w_{0y} = 2.44\frac{\lambda z_i}{s_y} \tag{6}$$

The minimum waist sizes corresponding to the maximum achievable resolutions of the central lenslet are $w_{0x}$ and $w_{0y}$. As expected, the maximum achievable resolutions are limited by the size of the lenslet along the $x$, and $y$ directions. This is the unwanted effect of the diffraction.

Without loosing the problem's generality, we assume that $s_x$ and $s_y$ are equal and therefore $w_x(z)$ and $w_y(z)$ are the same and named as $w(z)$. Similarly, $w_{0x}(z)$ and $w_{0y}(z)$ are equal and named as $w_0(z)$. The contribution of the central lenslet in the intensity of the visualized point source at the tilted plane can now be easily quantified by substituting the coordinates of the tilted plane in the intensity of the volumetric image formed by the central lenslet in the three-dimensional image space; i.e. $O_{m/2,n/2}(x,y,z)$:

$$O_{m/2,n/2}(x_\theta, y_\theta) = \frac{2P}{\pi w^2 (D - z_i + x_\theta \sin\theta_x + y_\theta \sin\theta_y)} e^{-\frac{-2(x_\theta^2 \cos^2\theta_x + y_\theta^2 \cos^2\theta_y)}{w^2(D - z_i + x_\theta \sin\theta_x + y_\theta \sin\theta_y)}} \tag{7}$$

In the same way, the contribution of the *pq*th lenslet of the array in the intensity of the visualized point source at the tilted plane can be calculated. It should be however noted that the Gaussian beam of the *pq*th lenslet is itself tilted. Since the tilt angle of the Gaussian beam with respect to the *x*, and *y*, coordinates are $\theta_x - \tan^{-1}\left(\frac{p.s_x}{D}\right)$, and $\theta_y - \tan^{-1}\left(\frac{q.s_y}{D}\right)$, respectively, the contribution of the *pq*th lenslet in the intensity of the visualized point source at the tilted plane is as follows:

$$O_{pq}(x_\theta, y_\theta) = \frac{d_{pq}^2}{(D+g)^2} \cdot \frac{w_0^2}{w^2(D - z_i + x_\theta \sin(\theta'_x) + y_\theta \sin(\theta'_y))} e^{-\frac{x_\theta^2 \cos^2\theta'_x + y_\theta^2 \cos^2\theta'_y}{w^2(D - z_i + x_\theta \sin\theta'_x + y_\theta \sin\theta'_y)}} \tag{8}$$

where

$$\theta'_x = \theta_x - \tan^{-1}\left(\frac{p.s_x}{D}\right)$$
$$\theta'_y = \theta_y - \tan^{-1}\left(\frac{q.s_y}{D}\right) \tag{9}$$

and

$$d_{pq} = \frac{D+g}{D}\sqrt{(ps_x)^2 + (qs_y)^2} \tag{10}$$

is the distance between the considered central image point and the corresponding pixel of the *p*,*q*th elemental image. Therefore, (8) could be simplified as follows:

$$O_{pq}(x_\theta, y_\theta) = \frac{\sqrt{(ps_x)^2 + (qs_y)^2}}{D^2} \cdot \frac{w_0^2}{w^2(D - z_i + x_\theta \sin(\theta'_x) + y_\theta \sin(\theta'_y))} e^{-\frac{x_\theta^2 \cos^2\theta'_x + y_\theta^2 \cos^2\theta'_y}{w^2(D - z_i + x_\theta \sin\theta'_x + y_\theta \sin\theta'_y)}} \tag{11}$$

Since there is an $m \times n$ lens-array in the system, the intensity of the visualized point source at the arbitrarily tilted plane; ($x_\theta y_\theta$),can be written as:

$$O(x_\theta, y_\theta) = \sum_{p=0}^{m-1} \sum_{q=0}^{n-1} O_{pq}(x_\theta, y_\theta) \tag{12}$$

Now that the image of the point source at the tilted plane is obtained, the radial extent of the optically visualized point source at the tilted plane can be written as follows:

$$w_{\theta_x,\theta_y} = \sqrt{\int_0^\infty \int_0^\infty \left(x_\theta^2 + y_\theta^2\right) O\left(x_\theta, y_\theta, D\right) dx_\theta dy_\theta} \tag{13}$$

where *O*(*x,y,D*) is the intensity of the reconstructed 3D image of the input point source in (*x,y,D*).The resolution of the system along $x_\theta$ and $y_\theta$ directions at *z*=*D* from the lens-array, where the optical point source was originally placed, is inversely proportional to the above mentioned radial extent of the optically visualized point source.

Further simplification of this expression shows that increasing the tilt angle decreases the overall resolution of the system. The field of view of the system can be defined as the maximum value of the tilt angle above which the resolution falls below a certain level. The best resolution is expected to be achieved when the point source is placed at the vicinity of the focused plane, i.e. $z_i$.

There are two different modes of imaging: one is the real/virtual mode with *g*>*f*/*g*<*f*, and the other is the focused mode with *g*= *f* [16], [21]. In the latter case, the focused plane $z_i$ tends to infinity and thus the extent of the beam does not strongly depend on the tilt angles. Although the overall resolution of the focused imaging mode is rather low, its field of view is quite large.

These facts are demonstrated via a numerical example. Consider a typical InI system with a $16 \times 16$ lens-array and with $s_x$ = $s_y$= 10*mm*. Two different imaging modes are considered. First, the real/virtual imaging mode with *g*=50 *mm*, and *f*=35*mm* is considered ($z_i$ =360*mm*). The radial extent of the visualized point source is calculated versus the tilt angles for *D* = 360 *mm*, and *D*= 450 *mm*. The obtained results are shown in Fig. 2. The resolution of the former case with *D*= 360 *mm* is larger than the resolution of the latter case with *D*=450 *mm*. This is not surprising because the resolution is expected to be higher when it is calculated at the vicinity of the focused plane at $z$ = 360 *mm*. If the acceptable spot size (inverse of the resolution) should not exceed 1.5 times the minimum spot size which occurs at $z = z_i$, the acceptable viewing angle is between $-15^\circ$ to $15^\circ$ in both $x$ and $y$ directions. This value is the average viewing angle obtained from Fig. 2.a and 2.b. Finally, the focused imaging mode with $g = f$ = 35 *mm* is considered. Similarly, the radial extent of the visualized point source is calculated versus the tilt angles for $D$ =2 *m*, and $D$ = 6 *m*. The obtained results are shown in Fig. 3. The resolution is not sensitive to the tilt angle but is much lower than the resolution of the real/virtual imaging mode. As already mentioned, the focused plane is much farther than the place at which the resolution is calculated. Still, the resolution of the system with $D$ = 6 *m* is slightly better than the resolution of the system with $D$ = 2 *m*. This is because the resolution at farther points is expected to be higher in the focused imaging mode. Again, if the acceptable spot size (inverse of the resolution) should not exceed 1.5 times the minimum spot size, the acceptable range of viewing angles is around $50^\circ$ in both negative and positive $x$ and $y$ directions. This is shown in Fig.3.

## III. Free View Reconstruction: Formulation and Simulations

To demonstrate the meaningfulness of the results obtained in the previous section; here, the mathematical formulation for free view reconstruction of three-dimensional images including the effects of diffraction and defocus aberration are provided. Then, the formulation is employed to reconstruct tilted views of two typical three-dimensional scenes whose elemental images are directly captured via a typical InI system with a $16 \times 16$ lens array. It should be however mentioned that the considered values for the number of elemental images and the other parameters used in the capturing part of the experiment are different from a practical display device which certainly have a large number of lenslets. Our previous analysis however shows that these factors merely scale the spot size [12]. Therefore, the derived conclusions could be confidently generalized.

### *A. Mathematical Formulation*

Once again, it is assumed that the InI system is composed of an $m \times n$ lens-array placed at distance $g$ from an ideal display device whose resolution is high enough to not restrain the overall resolution of the system. Computational image reconstruction neglecting the effects of diffraction and defocus aberration in normal view direction at any arbitrary axial distance; $z$, has been already reported [22]. The intensity of the $pq$th elemental image, $I_{pq}$, is back projected to the reconstruction plane and thus the contribution of the $pq$th elemental image in the volumetric three-dimensional image, $O_{pq}(x,y,z)$, is written as [22]:

$$O_{pq}(x,y,z)=\frac{I_{pq}\left(s_x p-\dfrac{x-s_x p}{M}, s_y q-\dfrac{y-s_y q}{M}\right)}{(z+g)^2+\left[(x-s_x p_y)^2+(y-s_y q)^2\right](1+1/M)^2} \tag{14}$$

where $M=z/g$ is the magnification factor of each lenslet in the array.

The three-dimensional image in an arbitrarily tilted plane can be easily reconstructed by using the abovementioned expression when the effects of diffraction and defocus aberration are to be neglected. If the tilt angles of the tilted plane are $\theta_x$ and $\theta_y$ and if the center of the tilted plane is placed at distance $z=D$ from the center of the lens array on the longitudinal axis, the coordinates of the three-dimensional space can in accordance with (1) be written in terms of $D$, and the lateral coordinates of the tilted plane $x_\theta$, and $y_\theta$. Therefore, the contribution of the $pq$th elemental image in reconstruction of the three-dimensional image at the tilted plane is as follows:

$$\begin{aligned}O_{pq}(x_\theta,y_\theta)&=I_{pq}\left(s_x p-\frac{x_\theta\cos\theta_x-s_x p}{M_{x_\theta y_\theta}}, s_y q-\frac{y\cos\theta_y-s_y q}{M_{x_\theta y_\theta}}\right)/\\&\left((D+x_\theta\sin\theta_x+y_\theta\sin\theta_y+g)^2+\left[(x_\theta\cos\theta_x-s_x p_y)^2+(y_\theta\cos\theta_y-s_y q)^2\right](1+1/M_{x_\theta y_\theta})^2\right)\end{aligned} \tag{15}$$

where

$$M_{x_\theta y_\theta}=\frac{D+x_\theta\sin\theta_x+y_\theta\sin\theta_y}{g} \tag{16}$$

The denominator of (15) is the square of the distance from the pixel of elemental image $I_{pq}$ to the corresponding point of the inversely mapped elemental image at the desired tilted plane.

The effects of diffraction and defocus aberration can be easily taken into account by convolving the contribution of the $pq$th elemental image with the point-spread-function of the $pq$th lenslet:

$$\widetilde{O}_{pq}(x_\theta,y_\theta)=\int O_{pq}(u,v)\otimes\widetilde{P}_D(x_\theta,y_\theta,u,v)\,dudv \tag{17}$$

where the point-spread-function, $\widetilde{P}_D(.)$, is related to the Fourier transform of the pupil function of the lenslet; $P(u,v)$, [11]:

$$\widetilde{P}_D(x_\theta, y_\theta, u, v) = \left|\Im\left\{P(u,v) \times \exp\left[\frac{jk}{2}\left(\frac{1}{D + x_\theta \sin\theta_x + y_\theta \sin\theta_y} - \frac{1}{z_i}\right)(u^2 + v^2)\right]\right\}\right|^{\acute{e}} \quad (18)$$

Since there is an $m \times n$ lens-array in the system, the reconstructed image at the desired tilted plane can be written as:

$$\widetilde{O}(x_\theta, y_\theta) = \sum_{p=0}^{m-1}\sum_{q=0}^{n-1} \widetilde{O}_{pq}(x_\theta, y_\theta) \quad (19)$$

As expected, setting $\theta_y = \theta_x = 0$ and $\widetilde{P}_D(x_\theta, y_\theta, u, v) = \delta(x_\theta - u, y_\theta - v)$ results in normal view reconstruction of the three-dimensional image at distance $D$ from the lens array when the effects of diffraction and defocus aberration are neglected.

### B. Experimental Results

In a fashion similar to section II, the effects of diffraction and defocus aberration are studied for both real/virtual and focused imaging modes [23, 24]. In the real/virtual imaging mode ($g>f/g<f$), the 3D reconstructed image is expected to have a very good quality at the focused plane; $z_i$, where the defocus aberration is minimum. Increasing the tilting angle inflicts severe destructive effects and the field of view is rather limited. In the focused imaging mode ($g=f$); however, the field of view is not limited but the overall resolution is quite low. Since the elemental images are placed at the focal plane, each pixel is projected to a parallel bundle of rays whose width is as large as the pupil of lenslets in the array. Therefore, the resolution is expected to be comparable to the lens pitch and remains almost constant at different depths and viewpoints. All these facts, already justified via Gaussian beam approximation, are more rigorously verified in the following subsections.

#### a. Real/Virtual Imaging Mode

A three-dimensional scene made of three toys is imaged via a typical InI system in the real/virtual imaging mode. Instead of using a $16 \times 16$ array of square aperture lenses, 256 elemental images are recorded via a digital camera (Cannon EOS 40D) whose focal length is set to $f$ = 35*mm*.The imaging sensor of the camera is placed at $g$ = 50 *mm* away from the lens and therefore the scene is recorded via real/virtual imaging mode. The elemental images are recorded by displacing the camera to imitate the lens pitch of 10 *mm*. The full frame of the imaging sensor of the camera is used for recording each elemental image; therefore, the issue of unwanted interference between neighboring elemental images is resolved. There are $252 \times 378$ pixels in each elemental image. This experimental setup mimics an InI system with a $16 \times 16$ lens array similar to the one analyzed in the previous section.

Two different methods are employed to reconstruct the three-dimensional scene at different tilt angles. First, the rules of geometrical optics are followed and the contributions of all 256 elemental images (see (12)) are summed to reconstruct the original scene at $D$= 0.4 *m*, and tilt angles $\theta_x$= 0°,20°, 30°, 35°, 40°, and 50°. For simplicity's sake the imaging planes are just tilted in $x$ direction and thus $\theta_y$= 0°. The obtained results are shown in Fig. 4. This figure shows that the field of view of the system can be as high as 50° when the effects of diffraction are absent, i.e. when the scene is numerically reconstructed via geometrical optics.

To demonstrate the deteriorating effects of diffraction and defocus aberration, the second method for reconstruction of the scene considers the diffraction caused by the finite size of the camera lens. Calculating the point spread function of the camera lens whose aperture diameter is 10*mm*, the three-dimensional scene is reconstructed by using (18) at $D$ = 0.4*m* and tilt angles $\theta_x$ = 0°,10°, 12°, 15°, 17°, and 20°. The obtained results are shown in Fig. 5. As it was already discussed in the previous section, the field of view of the system is about 15° when the unwanted effects of diffraction are taken into account. It is worth noting that the best resolution is expected to be observed at the focused plane $z_i$= 0.36 *mm*. Given that $D$ = 0.4*m*, the reconstructed image has a better quality along the vertical line at $x_\theta = (D - z_i)/\cos\theta_x$.

#### b. Focused Imaging Mode

Another three-dimensional scene made of an optometrist toy is imaged via a typical InI system in the focused imaging mode. Once again, the same digital camera (Cannon EOS 40D) is employed to record the elemental images of the scene. The focal length of the camera lens is set to $f$ = 35 *mm*, and the imaging sensor of the camera is placed at the focal plane, i.e. at $g$ = 35 *mm* away from the lens. Like the previous section, the elemental images are recorded by displacing the camera to imitate the lens pitch of 10 *mm*. There are $36 \times 36$ pixels in each elemental image. Although it is possible to record 256 elemental images and mimic an InI system with a $16 \times 16$ lens array, recording more than 121 elemental images does not increase the resolution and is unnecessary.

Similarly, the unwanted effects of diffraction and defocus aberration on three-dimensional scene reconstruction at different tilted planes are studied. First, the rules of geometrical optics are followed and the scene is reconstructed at $D$ = 3 *m* and tilt angles $\theta_x$ = 0°, 20°, 30°, 40°, 45°, and 50°. The obtained results are shown in Fig. 6. The resolution is not very good even in the absence of unwanted diffraction effects. Second, the effects of unwanted diffraction are taken into account and the scene is reconstructed at $D$ = 3 *m* and tilt angles $\theta_x$ = 0°, 20°, 30°, 40°, 45°, and 50°. For simplicity's sake the imaging planes are just tilted in $x$ direction and thus $\theta_y$= 0°. According to Fig. 7 that shows the obtained results, the resolution is not much worse than the idealized case, where the effects of diffraction had been neglected.

These figures show that the focused imaging mode is robust to unwanted diffraction effects.

## IV. Conclusion

In this manuscript, the resolution and field of view of three-dimensional InI systems were estimated by using Gaussian beam approximation to consider the unwanted effects of diffraction. It was shown that the resolution of the real/virtual imaging mode is high yet sensitive to the viewing angle, while the resolution of the focused imaging mode is low yet robust to the changes in the viewing angle.

The mathematical formulation for numerical reconstruction of three-dimensional images at arbitrarily tilted planes was also provided. The formulation considers the finite size of the aperture of lenslets in the lens array and thus considers the unwanted effects of diffraction. To verify the results obtained by using the Gaussian wave approximations, two different three-dimensional scenes were recorded in both the real/virtual and the focused imaging modes. Using the presented mathematical formulation, the three-dimensional scenes were reconstructed at different viewing angles. As expected, the resolution of the real/virtual imaging mode was higher. Still, the resolution was severely deteriorated by the unwanted diffraction effects particularly at tilted planes with larger tilt angles. It was shown that the field of view of the real/virtual imaging mode is strongly limited by the diffraction effects. In contrast, the resolution of the focused imaging mode; despite being rather low, was very much resilient against the unwanted diffraction effects.

## Acknowledgment

First set of elemental images recorded in the real/virtual mode has been provided by Dr. Saeed Bagheri in the University of Connecticut. The second set of elemental images recorded in the focused mode has been taken by one us, Zahra Kavehvash, in the laboratory Prof. M. Martinez-Corral in the University of Valencia. The authors sincerely thank their help and acknowledge their contribution.

**Zhila Esna Ashari Esfahani** received the B.S. degree in electrical engineering from University of Tehran, Tehran, Iran, in 2009 and the M.Sc. degree in electrical engineering from Sharif University of Technology, Tehran, Iran, in 2012.

Her research interests are optics, 3D imaging, integral imaging and computational photography.

**Zahra Kavehvash** received the B.Sc. degree in electrical engineering and M.Sc. and Ph.D. degrees from Sharif University of Technology (SUT), Tehran, Iran, in 2005, 2007, and 2012 respectively.

Currently she is an assistant professor at the electrical engineering department, SUT, Tehran, Iran. Her research interests are optical 3D imaging, integral imaging, holography, optical signal processing, optical computing and optical communications.

**Khashayar Mehrany** was born in Tehran, Iran, on September 16, 1977. He received the B.Sc., M.Sc. and Ph.D., (*magna cum laude*) degrees from Sharif University of Technology, Tehran, Iran, in 1999, 2001, and 2005 respectively all in electrical engineering.

Since then has been with the Department of Electrical Engineering, Sharif University of Technology, where he is now an Associate Professor. His research interests include photonics and numerical treatment of electromagnetic problems.

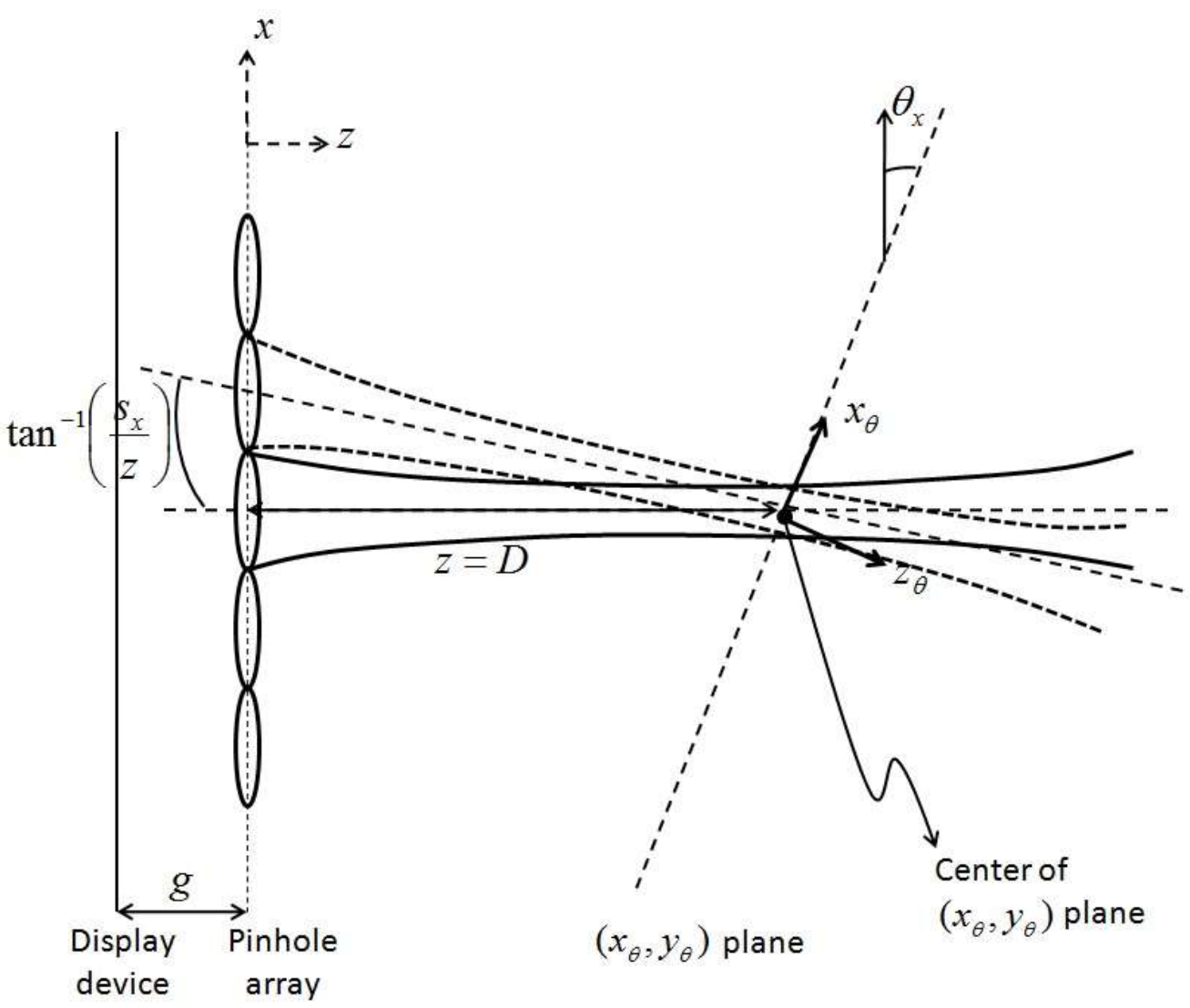

Fig. 1. Schematic structure for free-view reconstruction in $(x_\theta, y_\theta)$ plane shown in $x_\theta$ direction.

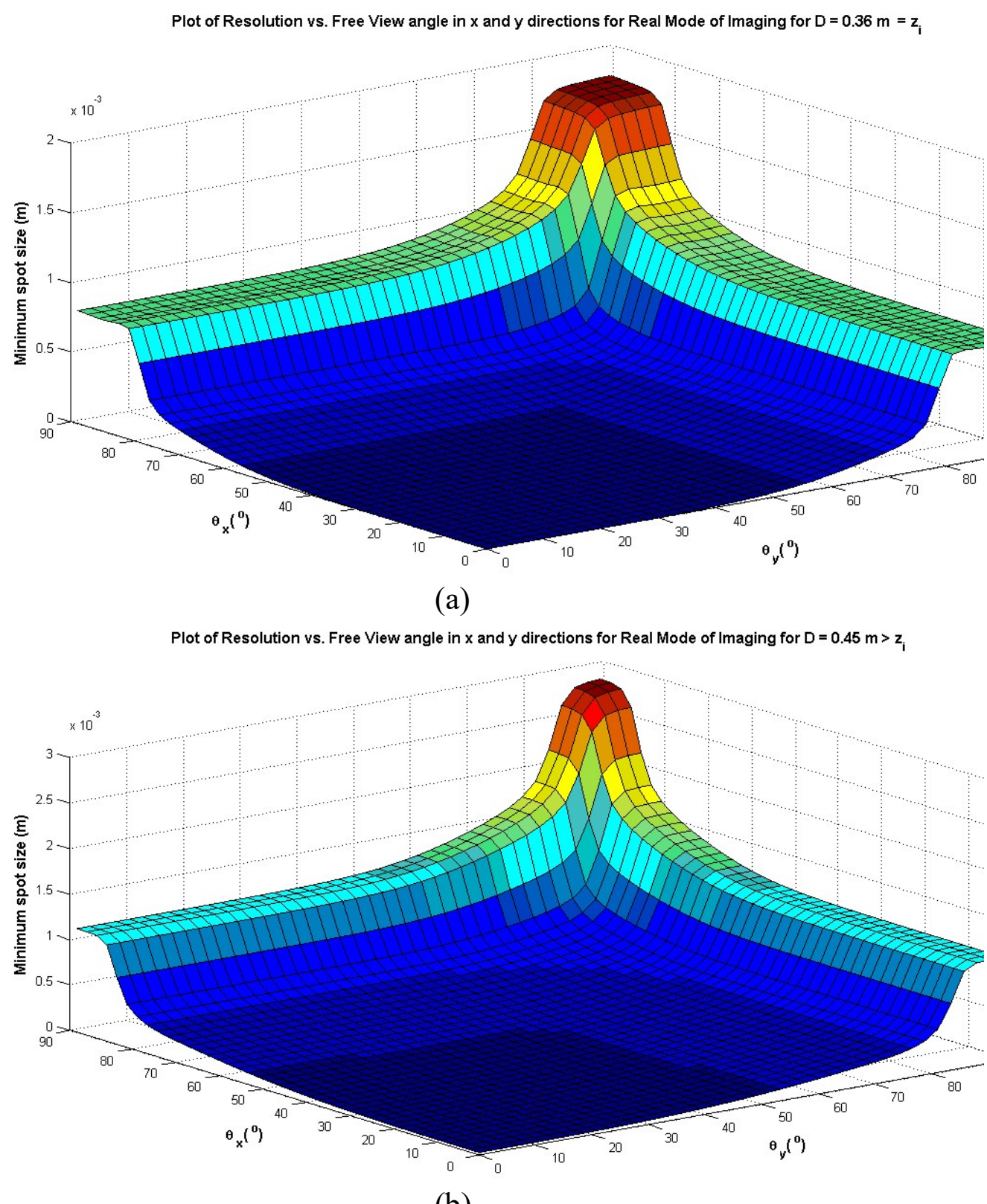

(a)

(b)

Fig. 2. The radial extent of the visualized point source versus the tilt angle in a real/virtual imaging mode with $z_i$ = *360 mm* when (a) *D* = *360 mm*, (b) *D* = *450 mm*.

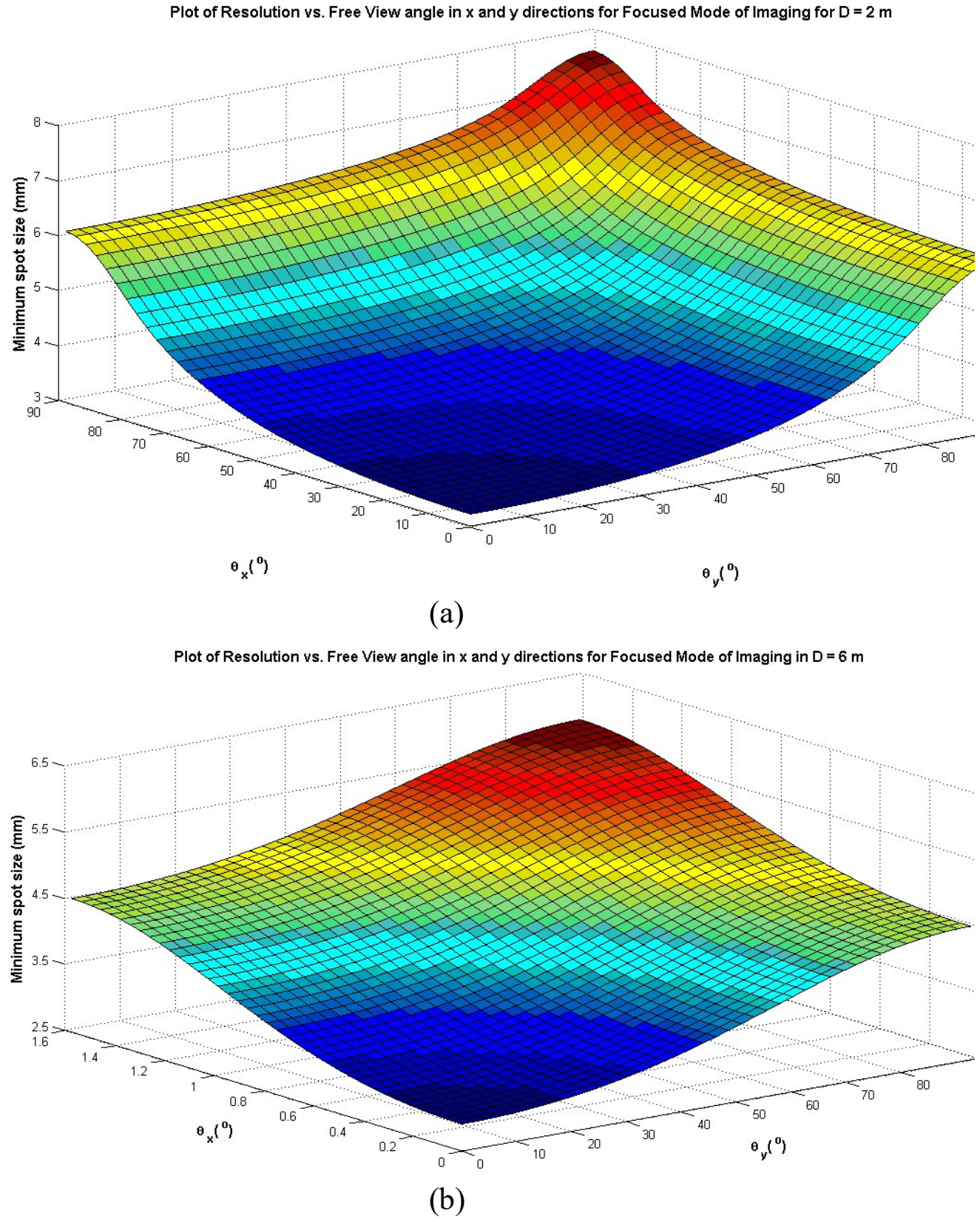

(a)

(b)

Fig. 3. The radial extent of the visualized point source versus the tilt angle in a focused imaging mode when (a) *D* = *2 m*, (b) *D* = *6 m*.

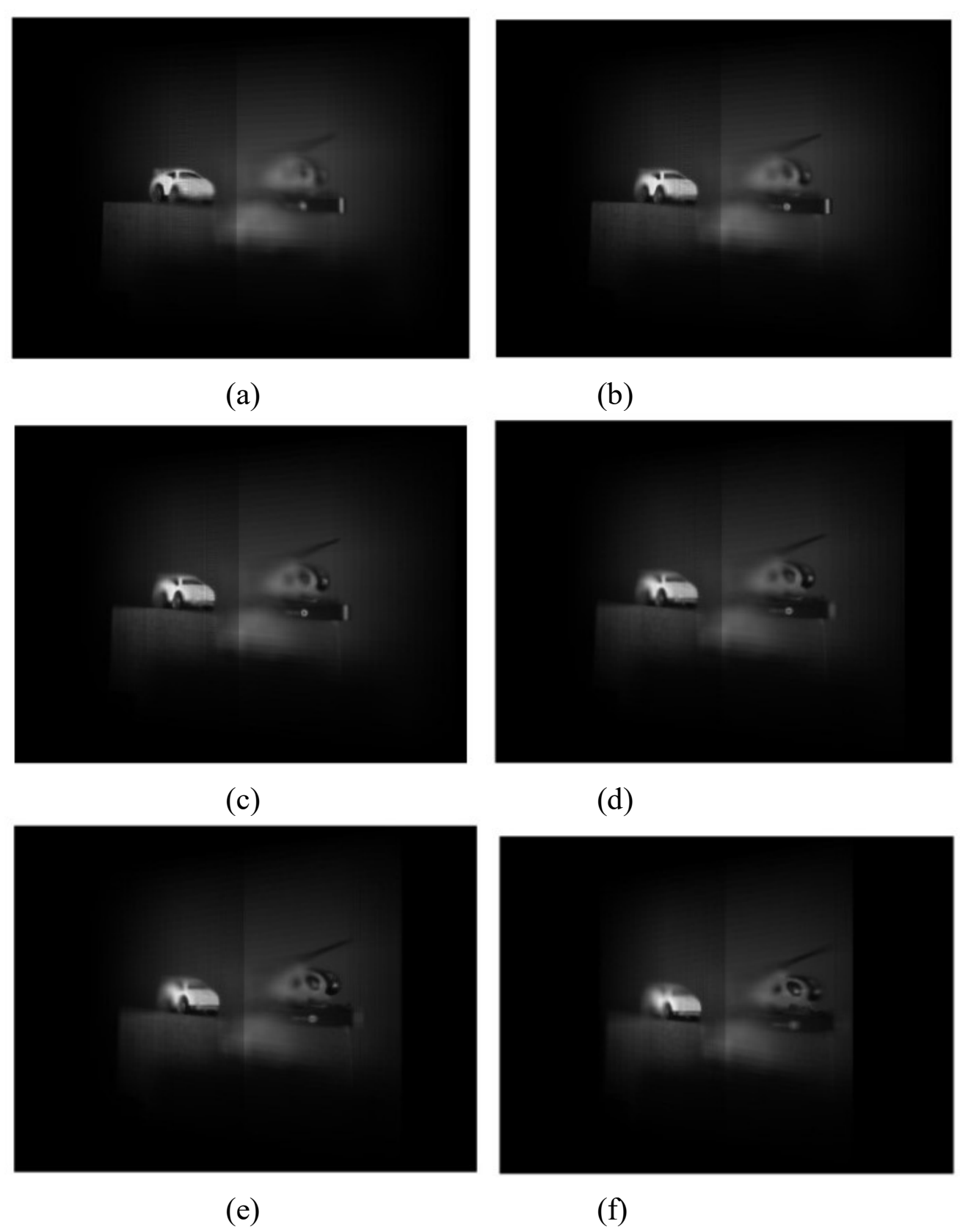

(a) (b)

(c) (d)

(e) (f)

Fig. 4. The reconstructed scene in real/virtual imaging mode when the unwanted effects of diffraction are neglected. The scene is reconstructed at tilted planes with tilt angle (a) 0°, (b) 20°, (c) 30°, (d) 35°, (e) 40°, (f) 50°.

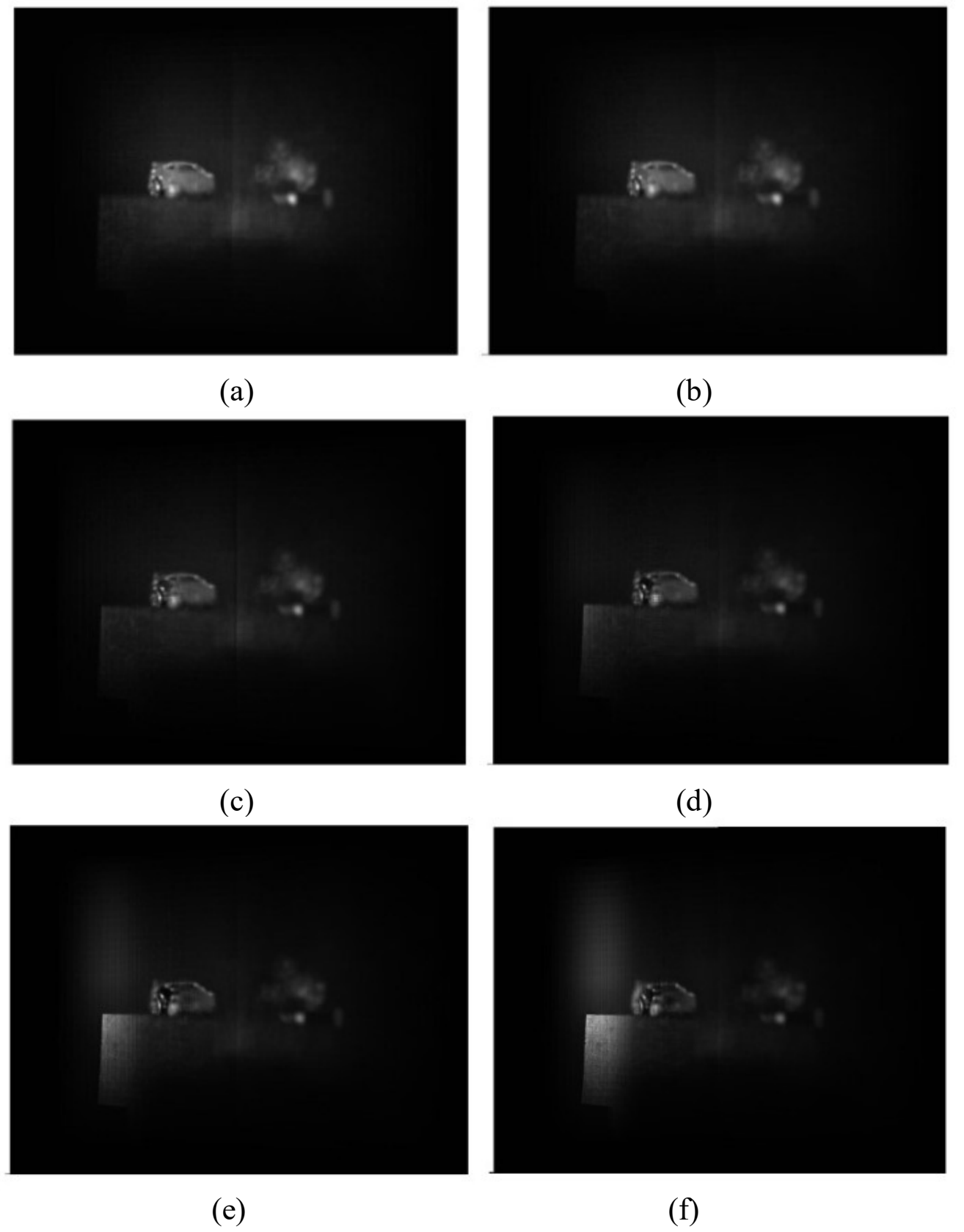

Fig. 5. The reconstructed scene in real/virtual imaging mode when the unwanted effects of diffraction are included. The scene is reconstructed at tilted planes with tilt angle (a) 0º, (b) 10º, (c) 12º, (d) 15º, (e) 17º, (f) 20º.

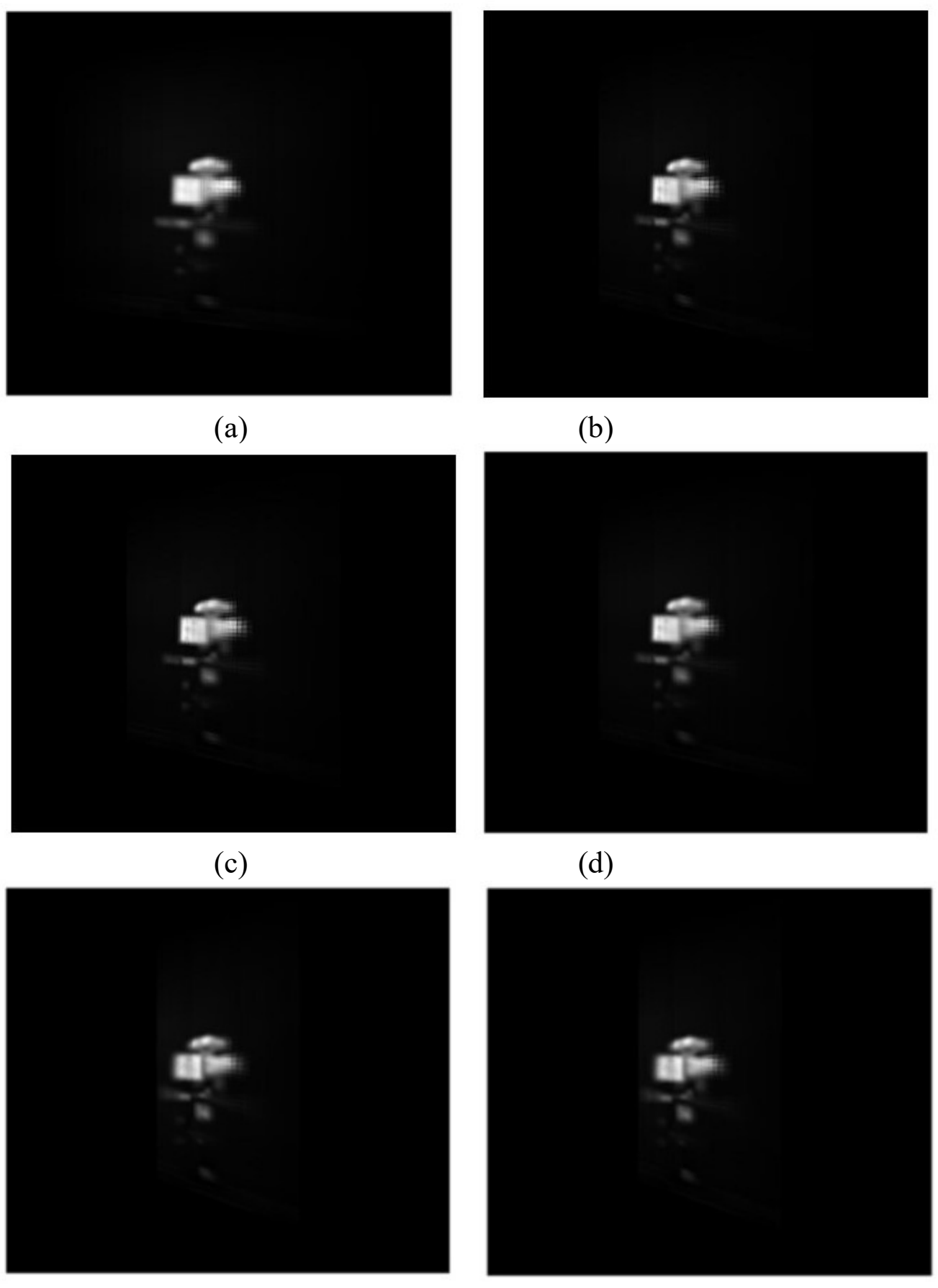

Fig. 6. The reconstructed scene in focused imaging mode when the unwanted effects of diffraction are neglected. The scene is reconstructed at tilted planes with tilt angle (a) 0º, (b) 20º, (c) 30º, (d) 40º, (e) 45º, (f) 50º.

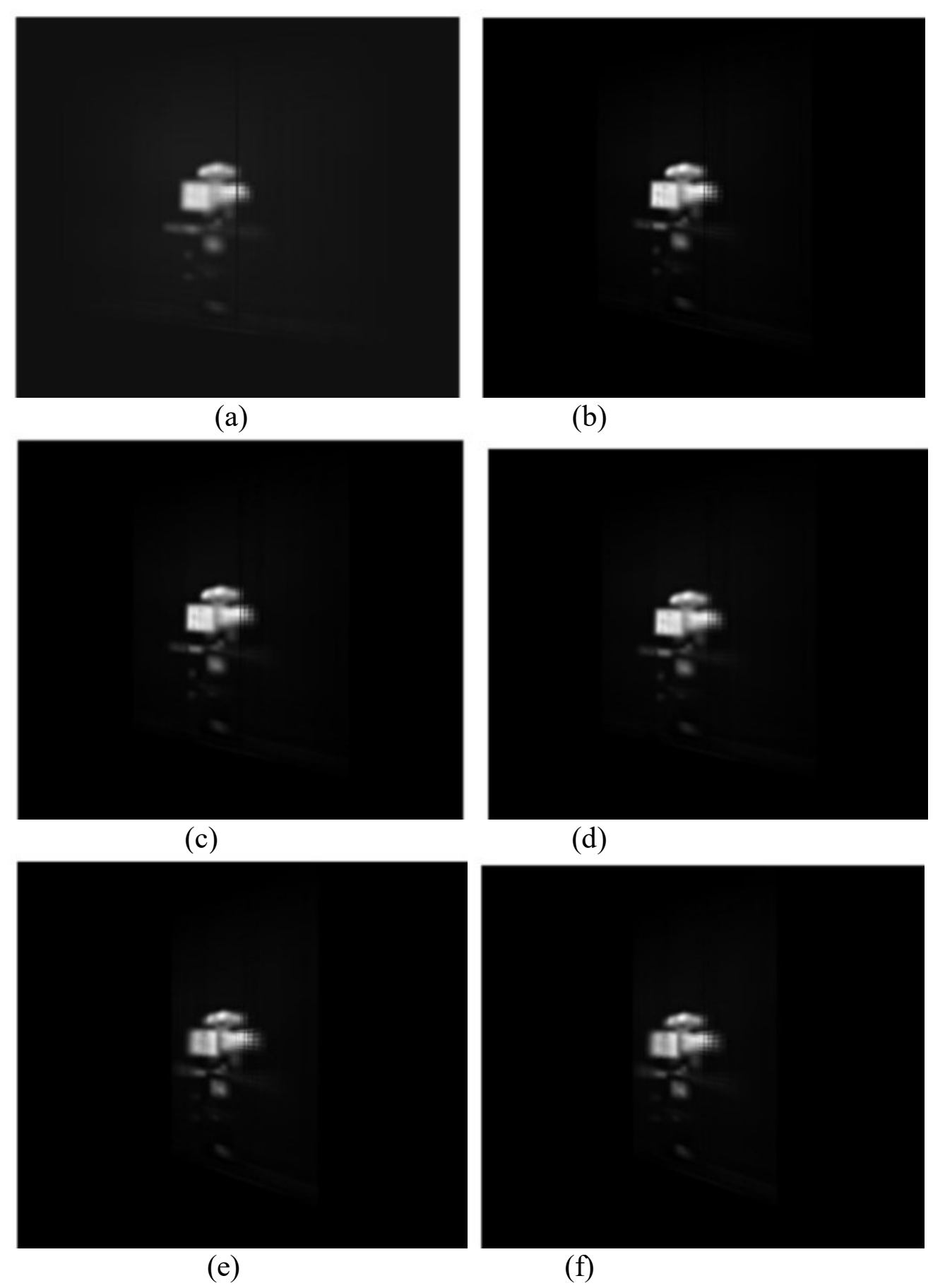

Fig. 7. The reconstructed scene in focused imaging mode when the unwanted effects of diffraction are included. The scene is reconstructed at tilted planes with tilt angle (a) 0º, (b) 20º, (c) 30º, (d) 40º, (e) 45º, (f) 50º.